\documentclass[preprint,12pt]{elsarticle}

\usepackage{amssymb}
\usepackage{graphicx}
\usepackage{subfigure}
\usepackage{ulem}  
\usepackage{color} 

\begin{document}

\begin{frontmatter}



\title{Nonlinear dynamic analysis of an optimal particle damper}


\author[1,2]{Mart\'{\i}n S\'anchez\corref{cor}}\ead{sanchez.martin@frlp.utn.edu.ar}
\author[1,3]{C. Manuel Carlevaro}
\cortext[cor]{Corresponding author. Telfax.: +54-221-482-4855}

\address[1]{Instituto de F\'{\i}sica de L\'{\i}quidos y Sistemas Biol\'ogicos (CONICET La Plata, UNLP), Calle 59 Nro 789, B1900BTE La Plata, Argentina.}
\address[2]{Centro de Ensayos Estructurales, Facultad Regional Delta, Universidad Tecnol\'ogica Nacional, Av. San Martín 1171, B2804GBW Campana, Argentina.}
\address[3]{U.D.B. F\'isica, Facultad Regional Buenos Aires, Universidad Tecnol\'ogica Nacional, Mozart 2300, C1407IVT Buenos Aires, Argentina.}

\begin{abstract}
We study the dynamical behavior of a single degree of freedom mechanical system with a particle damper. The particle (granular) damping was optimized for the primary system operating condition by using an appropriate gap size for a prismatic enclosure.  The particles absorb the kinetic energy of the vibrating structure and convert it into heat through the inelastic collisions and friction. This results in a highly nonlinear mechanical system.
Considering linear signal analysis, state space reconstruction, Poincar\'e sections and the determination of
maximal Lyapunov exponents, the motion of the granular system inside the enclosure is characterized for a wide frequency range. With the excitation frequency as control parameter, either regular and chaotic motion of the granular bed are found and their influence on the damping is analyzed.
\end{abstract}


\end{frontmatter}


\section{Introduction}
\label{intro}
In recent years, particle dampers (PD) have been studied extensively for use in harsh environments where other types of damping devices are not efficient. A PD is an element that increases the structural damping by inserting dissipative grains within holes in a vibrating structure or in a box attached to the primary system \cite{Panossian}.
The granular particles absorb the kinetic energy of the primary system and convert it into heat through inelastic collisions and friction between the particles and between the particles and the walls of the enclosure.
PD are effective over a wide frequency range. Moreover, PD are durable, inexpensive, easy to maintain and have a great potential for vibration and noise suppression for many applications (see \cite{Simonian} and \cite{XU}).

An appropriate treatment of the PD in a given structure requires careful analysis and design, parameters such as the size and shape of the particles, density, coefficient of restitution, size and shape of the enclosure, and the type of excitation of the primary system, among many other factors, are important in damping performance (see e.g. \cite{Marhadi} and \cite{Lu}). Due to its complexity, the understanding of damping mechanisms involved in PD is still limited.

The effectiveness of a PD is directly related to cooperative movements of the particles inside the cavity. Due to the inelastic collisions and friction, the system has a behavior highly nonlinear. The fundamental characteristic that affects the performance of the PD is the state of the granular media within the PD. Solid-, liquid- and gas-like phases have been identified for vibrating granular systems \cite{Saluena}.

Many authors have focused on studying the behavior of the damping device regardless of the complex internal dynamics of the particles. Therefore the theoretical models derived from single particle systems \cite{Friend,Duncan} are not applicable to analyze the nonlinear dynamical multi-particle systems. 

In this work, we use the Discrete Element Method (DEM) \cite{Cundall} to simulate the internal dynamics of a PD in order to analyze the nonlinear behavior and to characterize the granular system inside the cavity. These particle dynamic simulations have been used by others for investigating the behavior of these complex granular systems \cite{Mao,Saeki,Bai,Fang}. The method proposed by Sanchez and Pugnaloni \cite{Sanchez} to obtain the optimal gap size (space left between the granular layer and the top of the enclosure) through the particles effective mass is used to achieve the design with the best damping performance for the given conditions.

With the data obtained through simulations, the damping performance for a wide range of frequencies is investigated. We distinguish periodic and chaotic motion of the granular bed inside the enclosure at different excitation frequencies. Results show that the particles run from a periodic motion to a chaotic motion. Furthermore, we found that the optimal damping performance occurs in a window of periodicity. 

\section{The mechanical model}
\label{simul}
\begin{figure}[htp]
\begin{center}
\includegraphics[width=12cm]{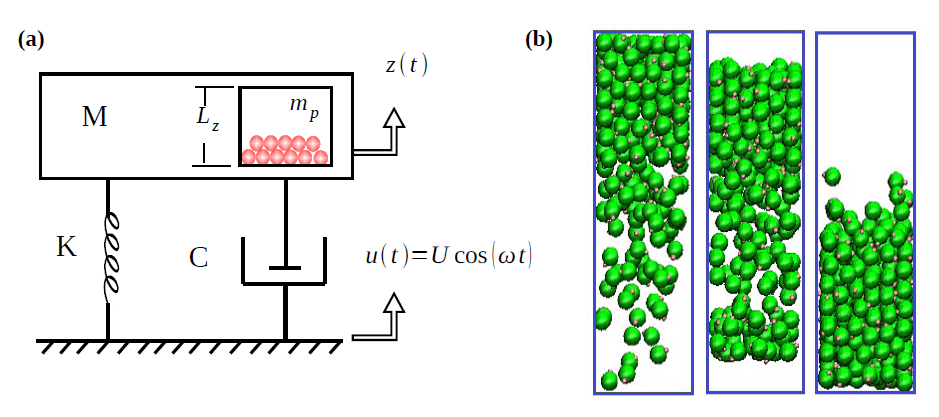}
\end{center}
\caption{\textbf{(a)} Model of the SDoF system with a particle damper. \textbf{(b)} Snapshots of the particles in the enclosure during a typical simulation.}
\label{fg:Fig. 1}
\end{figure}

The single degree of freedom (SDoF) primary system (see Fig.~\ref{fg:Fig. 1}) has been modeled with a mass $M = 2.37$ kg, a linear spring $K = 21500$ Nm$^{-1}$ and a viscous damper with damping constant $C = 7.6$ Nsm$^{-1}$. The PD consists of $N = 250$ spherical grains  in a prismatic enclosure made by six flat walls with the same material properties as the particles. The material properties of the particles (and walls) and the simulation parameters are listed in Table~\ref{tab:Tabla1}.

The gravitational field $g = 9.8$ ms$^{-2}$ is considered in the negative vertical direction. Although the SDoF system can only move in the vertical direction, the particles move freely inside the enclosure. 

The particles-particle interaction force $F_\mathrm{n}$ in the normal direction is based on the Hertz--Kuwabara--Kono model \cite{Schafer,Kruggel1}.
\begin{equation}
F_\mathrm{n} = -k_\mathrm{n}\alpha^{3/2}-\gamma_\mathrm{n}\upsilon_\mathrm{n}\sqrt{\alpha};
\label{normal}
\end{equation}
where $k_\mathrm{n}=\frac{2}{3}E\sqrt{\frac{R}{2}}(1-\upsilon^{2})^{-1}$ is the normal stiffness (with $E$ the Young's modulus, $\upsilon$ the Poisson's ratio and $R^{-1}=R_i^{-1}+R_j^{-1}$ and $R_i$ the radius of particle $i$), $\gamma_\mathrm{n}$ the normal damping coefficient of the viscoelastic contact, and $\upsilon_\mathrm{n}$ the relative normal velocity.

The particle-particle tangential force $F_\mathrm{s}$, is based on Coulomb's law of friction \cite{Schafer,Kruggel2}. We used a simplified model in which the friction force takes the minimum value between the shear damping force and the dynamic friction \cite{Schafer,Kruggel2}.
\begin{equation}
F_\mathrm{s} = -\min\left(\left|\gamma_\mathrm{s}\upsilon_\mathrm{s}\sqrt{\alpha}\right|,\left|\mu_\mathrm{d}F_\mathrm{n}\right|\right)\rm{sgn}\left(\upsilon_\mathrm{s}\right);
\label{tangential}
\end{equation}
where $\gamma_\mathrm{s}$ is the shear damping coefficient, $\upsilon_\mathrm{s}$ the relative tangential velocity between the two spheres in contact and $\mu_\mathrm{d}$ the dynamic friction coefficient. The sign function indicates that the friction force always opposes the direction of the relative tangential velocity.

The system is excited by the harmonic displacement of the base to which the spring and viscous damper are attached (see Fig.~\ref{fg:Fig. 1}). Let $u(t)$ and $z(t)$ be the displacement of the base and the primary mass, respectively. Then, the equation of motion for the system is given by:
\begin{equation}
M\ddot{z}(t) + C \dot{z}(t) + K {z}(t) = C \dot{u}(t) + K u(t) + F_\mathrm{part}(t) ; \label{ec2} \ \ \ \ \ \  u(t) = U\cos(\omega t)
\end{equation}
where $F_\mathrm{part}(t)$ is the $z$-component of the force resulting from all the interactions (normal and tangential) of the enclosure walls with the particles. The amplitude, $U$, and the angular frequency, $\omega$, of the harmonic vibrating base, are control parameters.
\begin{table}[htb]
\centering
\begin{tabular}{|c|c|}
\hline Property & Value \\
\hline
\hline
Young's modulus $E$& $2.03\times10^{11}$ Nm$^{-2}$ \\
\hline
Density & 8030 kgm$^{-3}$ \\
\hline
Poisson's ratio $\upsilon$ & 0.28 \\
\hline
Friction coefficient $\mu_\mathrm{d}$ & 0.3 \\
\hline
Normal damping coefficient $\gamma_\mathrm{n}$ & $3.660\times10^{3}$ kgs$^{-1}$m$^{-1/2}$ \\
\hline
Shear damping coefficient $\gamma_\mathrm{s}$ & $1.098\times10^{4}$ kg$s^{-1}$m$^{-1/2}$ \\
\hline
Excitation amplitude $U$ & 0.0045 m \\
\hline
Time step $\delta t$ & $8.75\times10^{-8}$ s \\
\hline
Time of simulation & 196.83 s \\
\hline
Particle radius & 0.003 m \\
\hline
Total particle mass $m_\mathrm{p}$ & 0.227 kg \\ 
\hline
Box lateral side $L_x$ and $L_y$ & 0.03675 m \\
\hline
Box height $L_z$ & 0.1225 m \\
\hline
Excitation frequency range & $0.5$ to $30.0$ Hz \\
\hline
\end{tabular}
\caption{Material properties of the particles and simulation parameters.}
\label{tab:Tabla1}
\end{table}

\section{Nonlinear dynamic analysis}
\label{NDA}
The mechanical model described in Section \ref{simul} can be numerically solved to obtain a scalar sequence of the position along the $z$-axis of the center of mass of the granular bed and the $z$-position of the box. It has been shown in experiments (see e.g. \cite{Bannerman}) that the motion of the granulate is well described by the motion of the center of mass, therefore, time series associated with the PD are available and these are valid to investigate the granular behavior inside the cavity and analyze the damping performance. 

The analysis of nonlinear dynamical systems from time series involves linear signal analysis to identify periodic motions, state space reconstruction, Poincar\'e sections and the determination of the MLE to identify chaotic behavior. We have used for this analysis the Tisean Software package \cite{Hegger}.

The nonlinear analysis requires several inputs for the reconstruction of the state space and the determination of the MLE. In this contribution the false nearest neighbors method has been used \cite{Kennel} to determine the minimal sufficient embedding dimension and the time delayed mutual information for the delay \cite{Fraser}. The algorithm proposed by Kantz \cite{Kantz} is used to obtain the MLE.

\section{Numerical results}
\label{res}
As shown in Table~\ref{tab:Tabla1}, we have simulated the vibration of the system for $196.83$ s. After an initial transient, the system reaches a steady state. A simple evaluation of the effective damping yield by the particles can be done by fitting the frequency response function (FRF) to a SDoF system including only a viscous damper. The amplitude of the response $X$ of a SDoF system with no PD is given by
\begin{equation}
X = U\left[\frac{K^2+(C_\mathrm{eff}\omega)^2}{(K-M_\mathrm{eff}\omega^2)^2+(C_\mathrm{eff}\omega)^2}\right]^{1/2};
\label{ecsol2}
\end{equation}

\begin{figure}[htp]
\begin{center}
\includegraphics[width=12cm]{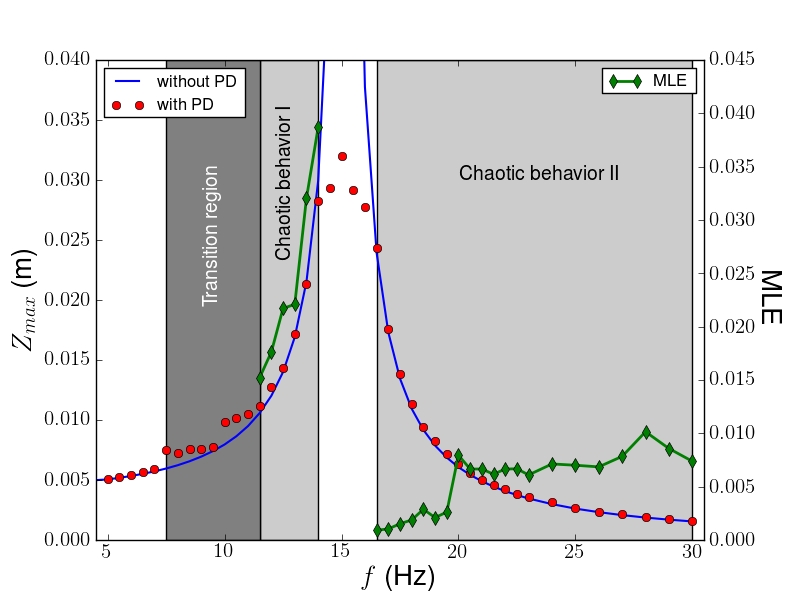}
\end{center}
\caption{Frequency response function of the primary system. The blue continuous line corresponds to the SDoF system without particles in the enclosure and with a mass equivalent to the system with the PD. The red circles correspond to the FRF for the optimum PD. The green diamonds are the MLE for each frequency in the chaotic region. The dark gray area corresponds to the transition to chaos, the gray area corresponds to the chaotic zone, and the white area to the periodic motion of the trajectory of the center of the mass of the granular bed.}
\label{fg:frf}
\end{figure}

We carry out a least-squares curve fitting of the DEM data with Eq.~\ref{ecsol2}. The values of $K$, $U$ and $C$ are fixed to the corresponding values in our simulations and $M_\mathrm{eff}$ is the fitting parameter. Figure \ref{fg:frf} presents the cavity position as a function of excitation frequency compared with the fitting curve without PD.

Several works \cite{Saeki,Sanchez2} predicted a optimal gap value yield the best damping performance. We use the particles effective mass to obtain the optimal gap size \cite{Sanchez}. The granular bed at rest has an approximate height of $L_z = 0.039$ m and the gap ($L_z - 0.039$ m). The FRF is complex for these conditions \cite{Liu}.

\subsection{Periodic behavior}
\label{perio}
\begin{figure}[htp]
\begin{center}
\includegraphics[width=4cm, clip]{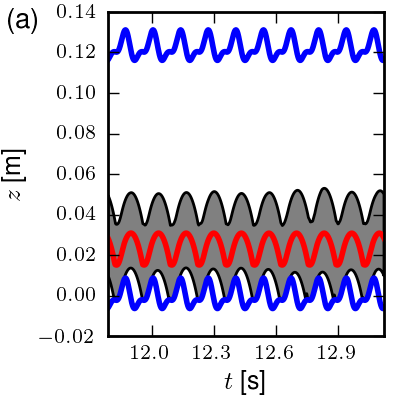}
\includegraphics[width=4cm, clip]{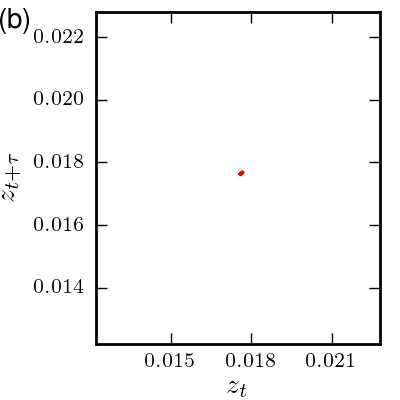}
\includegraphics[width=4cm, clip]{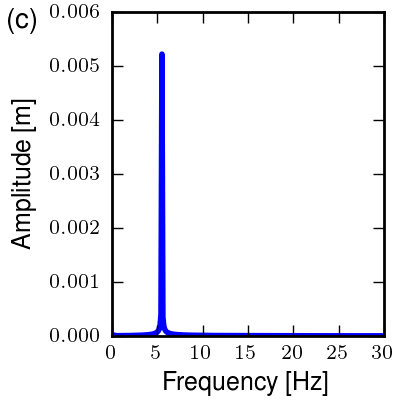}
\includegraphics[width=4cm, clip]{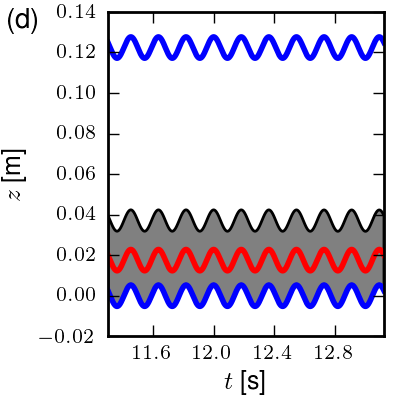}
\includegraphics[width=4cm, clip]{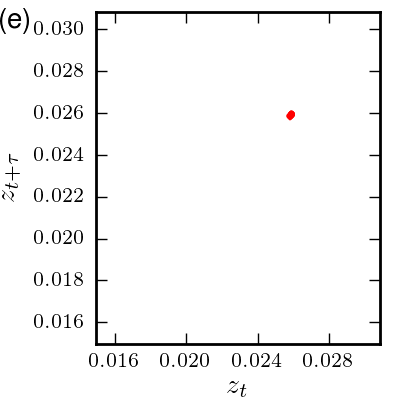}
\includegraphics[width=4cm, clip]{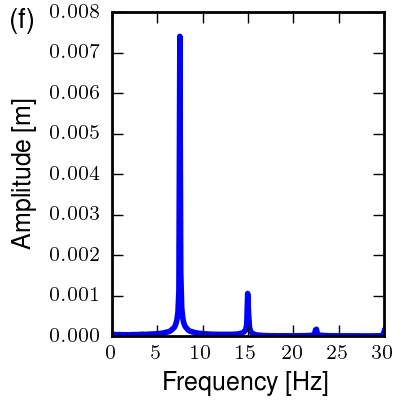}
\end{center}
\caption{Granular bed response at $f=5.5$ Hz and at $f=7.5$ Hz. \textbf{(a)} Displacement of particles against the enclosure for $f=5.5$ Hz. The solid lines show the position of floor and ceiling of the enclosure and the colored area indicates the limits of the granular bed defined as the position of the uppermost and lowermost particle. The red continuous lines is the motion of the center of mass of the grains. \textbf{(b)} Poincar\'e map for $f=5.5$ Hz. \textbf{(c)} FFT of the trajectory of the center of the mass for $f=5.5$ Hz. \textbf{(d)} Displacement of particles against the enclosure for $f=7.5$ Hz (Colors as in Fig.~\ref{fig:fig3}(a)). \textbf{(e)} Poincar\'e map for $f=7.5$ Hz. \textbf{(f)} FFT of the trajectory of the center of the mass for $f=7.5$ Hz.}
\label{fig:fig3}
\end{figure}

At low excitation frequency (much lower than the resonant frequency), the acceleration of the primary system is less than the acceleration of gravity and the granular bed lies always in contact with the floor of the box, thus the center of the mass of the granular bed follows a regular movement. This behavior can been seen in Fig.~\ref{fig:fig3}.

In Fig.~\ref{fig:fig3}(a) we plot the trajectory along the $z$-axis of the center of mass of the particles and the motion of the floor and ceiling of the enclosure over a few periods of excitation in the steady state regime for an excitation frequency $f=5.5$ Hz. We have indicated the position of the granular bed inside the enclosure by a band limited by the $z$-coordinates of the uppermost and lowermost particle at any given time. To show the regular motion of the center of mass at this frequency, we represent in Fig.~\ref{fig:fig3}(b) the Poincar\'e map (through a stroboscopic map), in this figure can been seen a classical Poincar\'e section for a periodic motion of a time series. Moreover, the reconstruction of the attractor of the center of mass (see Fig.~\ref{fig:fig8}(a)) is typical for regular movement, in addition the FFT (Fig.~\ref{fig:fig3}(c)) shows a peak at the excitation frequency of the primary system. 

Clearly, as the response level is too low, there is almost no relative motion between the particles and between the particles and the enclosure. Therefore, the grains can indeed be viewed as a lumped mass resting on the enclosure floor, and the damping effect is very small. In Fig. \ref{fg:frf} can been seen that the response of the primary system at this low frequency is similar to the system without the PD. 

When the excitation frequency increases ($f=7.5$ Hz), the inertia effect of the particles exceeds the gravity force and the granular motion develops. In Fig.~\ref{fig:fig3}(d) we plot the motion of the granular bed. This figure shows that there is significant particle motion which also leads to impacts between the grains and the enclosure floor. The excitation level is not drastic yet and the particles do not have enough energy to impact the ceiling of the enclosure, then the center of mass continues with a periodic motion.

The Poincar\'e section (Fig.~\ref{fig:fig3}(e)) is analogous to the one observed for $f=5.5$ Hz. The signal has a harmonic spectrum containing only frequency components whose frequencies are multiples of the fundamental frequency (Fig.~\ref{fig:fig3}(f)).

\subsection{Transition region}
\label{trans}
With further increase in $f$ beyond $7.5$ Hz a transition phase begins, which ends around the frequency $f=11.0$ Hz. The granular bed switches between quasi-periodic motions and periodic motions. 
\begin{figure}[htp]
\begin{center}
\includegraphics[width=4cm, clip]{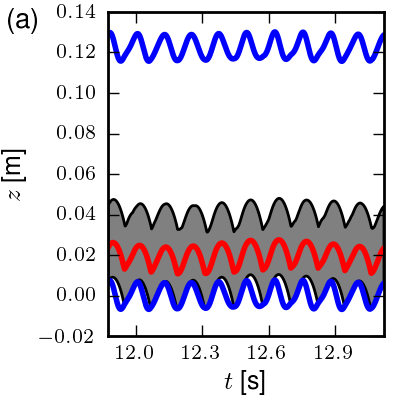}
\includegraphics[width=4cm, clip]{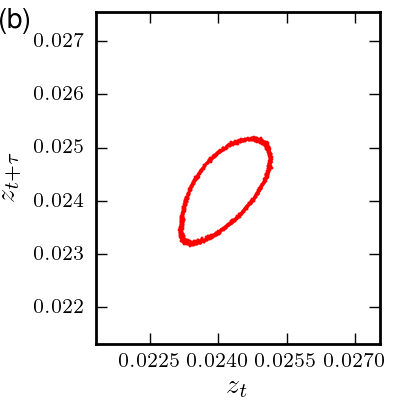}
\includegraphics[width=4cm, clip]{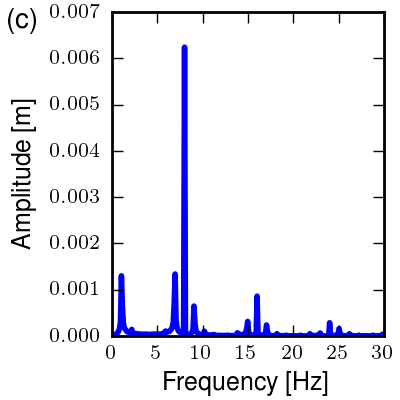}
\includegraphics[width=4cm, clip]{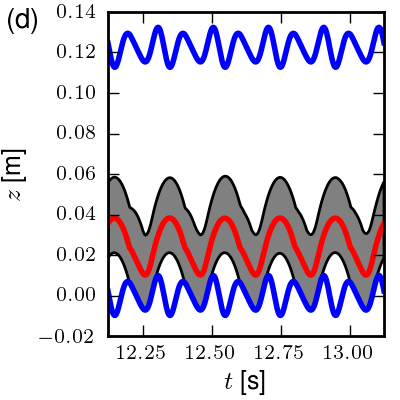}
\includegraphics[width=4cm, clip]{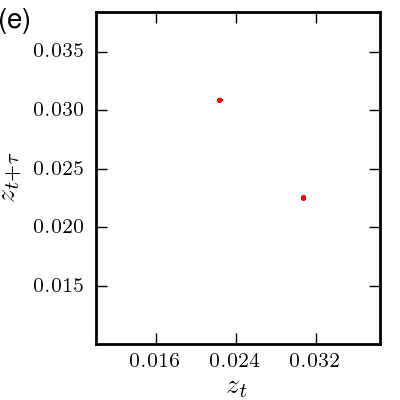}
\includegraphics[width=4cm, clip]{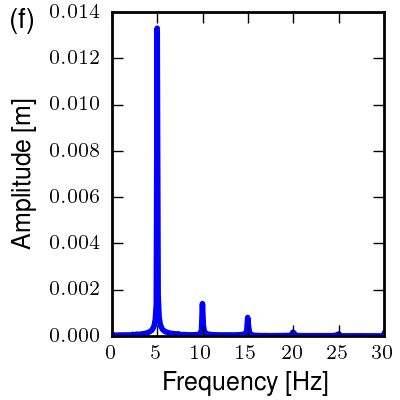}
\end{center}
\caption{Granular bed response at $f=8.0$ Hz and at $f=10.0$ Hz. \textbf{(a)} Displacement of particles against the enclosure at $f=8.0$ Hz (Colors as in Fig.~\ref{fig:fig3}(a)). \textbf{(b)} Poincar\'e map for $f=8.0$ Hz. \textbf{(c)} FFT of the trajectory of the center of the mass for $f=8.0$ Hz. \textbf{(d)} Displacement of particles against the enclosure at $f=10.0$ Hz (Colors as in Fig.~\ref{fig:fig3}(a)). \textbf{(e)} Poincar\'e map for $f=10.0$ Hz. \textbf{(f)} FFT of the trajectory of the center of the mass for $f=10.0$ Hz.}
\label{fig:fig4}
\end{figure}

For quasi-periodic motions we observe a torus attractor as shown in Fig.~\ref{fig:fig4}(b) for $f=8.0$ Hz. The large number of points that lay on a smooth closed curve in the Poincar\'e map confirms that the response of the granular bed is indeed quasi-periodic, where the response is made up of two frequencies of which the ratio of one of these frequencies to the other is an irrational number. The FFT shows a peak at the forcing frequency, and three other secondary peaks; of which one correspond to the particles response frequency and other two are simply the sum and difference of the forcing and response frequencies. The occurrence of these last peaks, that correspond to irrational fractions of the forcing frequency, is evidence that the granular bed response is quasi-periodic.

Alternating between these quasi-periodic motions we observed regular motions, which are characterized by clouds of points in the Poincar\'e sections and sub-harmonic peaks in the FFT. We show an example in Fig.~\ref{fig:fig4}(e), where we find a period-2 motion at $f=10.0$ Hz.

The energy of the particles inside the PD remains low in this region and no grains can reach the ceiling of the box. The increase in the amplitude of vibration of the primary system produces low granular densities upon each impact with the walls of the enclosure and the transfer of the momentum is not significant (see Fig.~\ref{fig:fig4}(a) and Fig.~\ref{fig:fig4}(d)), therefore the granular damping performance is poor and the response of the primary system follows with similar characteristic to the system without the PD (see Fig.\ref{fg:frf}).

\subsection{Chaotic behavior I}
\label{chaosI}
\begin{figure}[htp]
\begin{center}
\includegraphics[width=4cm, clip]{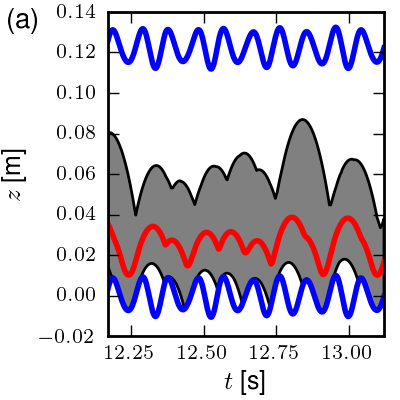}
\includegraphics[width=4cm, clip]{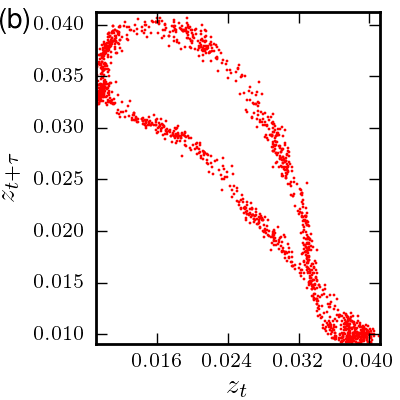}
\includegraphics[width=4cm, clip]{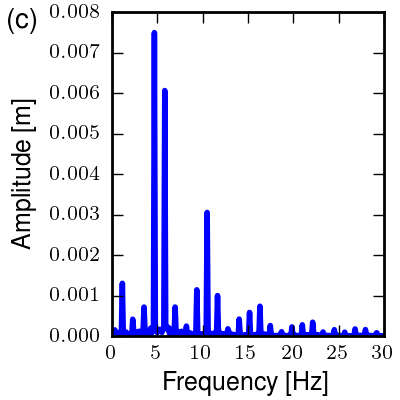}
\includegraphics[width=4cm, clip]{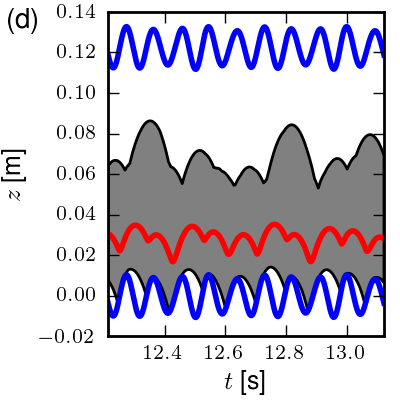}
\includegraphics[width=4cm, clip]{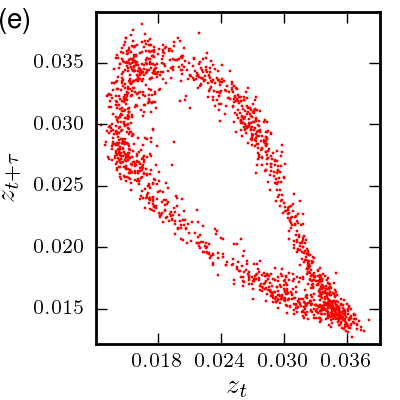}
\includegraphics[width=4cm, clip]{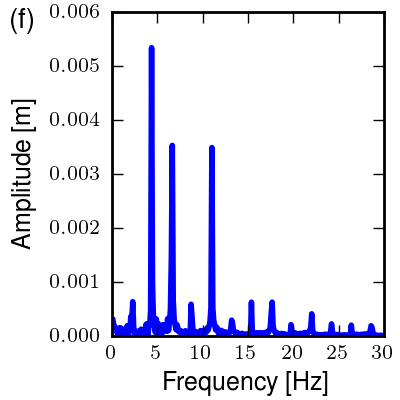}
\includegraphics[width=4cm, clip]{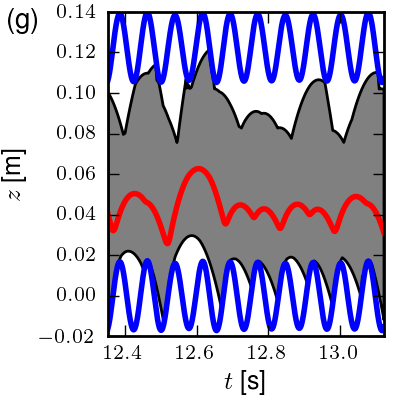}
\includegraphics[width=4cm, clip]{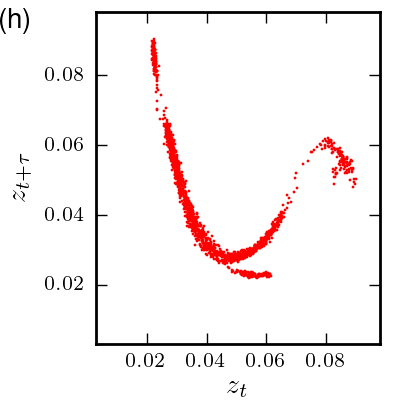}
\includegraphics[width=4cm, clip]{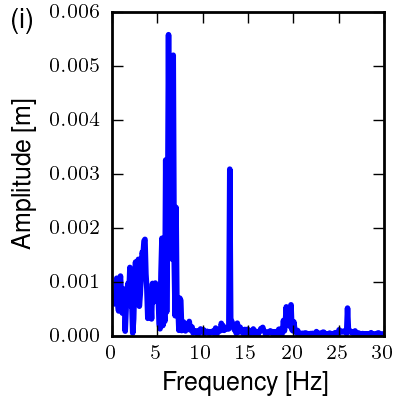}
\end{center}
\caption{Granular bed response at $f=10.5$, at $f=11.0$ Hz and at $f=13.0$ Hz. \textbf{(a)} Displacement of particles against the enclosure at $f=10.5$ Hz (Colors as in Fig.~\ref{fig:fig3}(a)). \textbf{(b)} Poincar\'e map for $f=10.5$ Hz. \textbf{(c)} FFT of the trajectory of the center of the mass for $f=10.5$ Hz. \textbf{(d)} Displacement of particles against the enclosure at $f=11.0$ Hz (Colors as in Fig.~\ref{fig:fig3}(a)). \textbf{(e)} Poincar\'e map for $f=11.0$ Hz. \textbf{(f)} FFT of the trajectory of the center of the mass for $f=11.0$ Hz. \textbf{(g)} Displacement of particles against the enclosure at $f=13.0$ Hz (Colors as in Fig.~\ref{fig:fig3}(a)). \textbf{(h)} Poincar\'e map for $f=13.0$ Hz. \textbf{(i)} FFT of the trajectory of the center of the mass for $f=13.0$ Hz.}
\label{fig:fig5}
\end{figure}

As the value of frequency is increased to $10.5$ Hz, wrinkling of the torus in the Poincar\'e section is observed. The distortion of the torus becomes more significant as $f$ is further increased resulting in the breaking up of the continuous curve, forming a fractal torus, at $f=11.0$ Hz. Fig.~\ref{fig:fig5} shown the behavior at $f=10.5$ Hz and at $f=11.0$ Hz.

The response level is high in the range from $f=11.0$ Hz to $f=14.0$ Hz, there is significant relative motion between the particles and between the particle and the enclosure with enough energy to impact the ceiling (see e.g. Fig.~\ref{fig:fig5}(g) for $f=13.0$ Hz). At these frequencies the Poincar\'e map, which consists of a large number of points lying on an open curve (Fig.~\ref{fig:fig5}(h)), suggest that the attractor of the grains response is strange. This in turn indicates that the response could possibly be chaotic.

The broadband frequency observed in the power spectrum plot (see Fig.~\ref{fig:fig5}(i)) suggests that the motion of particles reaches a chaotic regime. The MLE of the grains response for the range between $f=11.0$ Hz to $f=14.0$ Hz are shown in Fig.\ref{fg:frf}. These positive values confirm that the granular bed motion is indeed chaotic. Moreover, an strange attractor is found for these range of frequencies and we show it for $f=13.0$ Hz in (Fig.~\ref{fig:fig8}(b)).

\subsection{Optimal damping}
\label{opti}
\begin{figure}[htp]
\begin{center}
\includegraphics[width=4cm, clip]{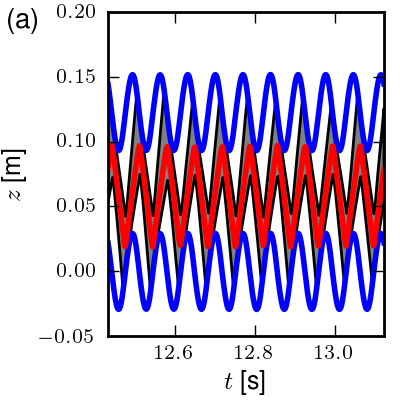}
\includegraphics[width=4cm, clip]{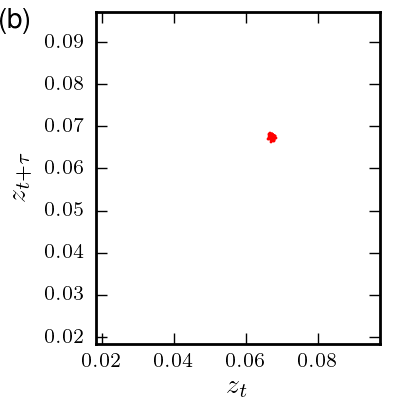}
\includegraphics[width=4cm, clip]{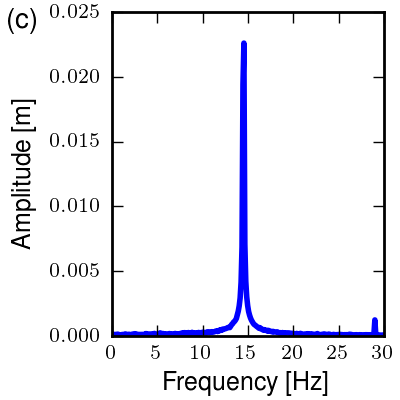}
\end{center}
\caption{Granular bed response at $f=14.5$ Hz. \textbf{(a)} Displacement of particles against the enclosure (Colors as in Fig.~\ref{fig:fig3}(a)). \textbf{(b)} Poincar\'e map. \textbf{(c)} FFT of the trajectory of the center of the mass.}
\label{fig:fig6}
\end{figure}

In the range between $f=14.5$ Hz to $f=16.0$ Hz, we found a window of periodicity, where the particles impact the enclosure two times per period of excitation. At these frequencies, the granular bed enters a solid-like state, colliding both the floor and the ceiling out of phase. We plot this behavior in Fig.~\ref{fig:fig6}. Note that the compactation of the granular bed is high in this regimen and the transfer of momentum at impact also increases.

The absolute velocities of the center of mass of the particles and the primary system are opposite to each other at the instant immediately before contact. Due to maximization of the relative velocity, the PD dissipates the maximum energy. As shown in Fig.\ref{fg:frf}, the maximum effective damping is obtained in this range. We show an example of the periodic attractor at $f=14.5$ Hz in (Fig.~\ref{fig:fig8}(c)).

\subsection{Chaotic behavior II}
\label{chaosII}
Further increase in the control parameter resulted in chaotic vibration in the response of the center of mass. This chaotic behavior of the granulate is observed from $f=16.5$ Hz immediately after the window of periodicity. At these frequencies (above the primary system resonant frequency) the motion of the grains is very intense and the particles enter a gas-like state. The excitation level is drastic and the granular bed expands significantly (very low compactation). Due to the high energy in the system, only a few particles hit the walls at each instant and again the systems response is similar to the system without the PD (see Fig.\ref{fg:frf}).
\begin{figure}[htp]
\begin{center}
\includegraphics[width=4cm, clip]{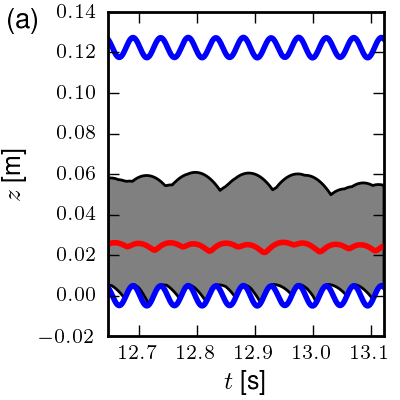}
\includegraphics[width=4cm, clip]{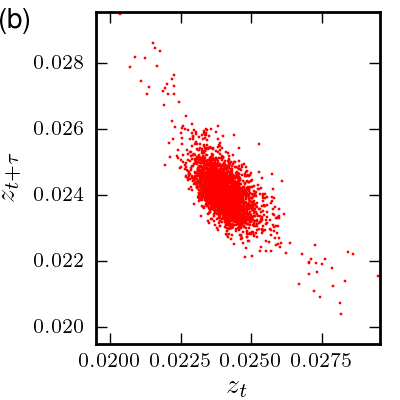}
\includegraphics[width=4cm, clip]{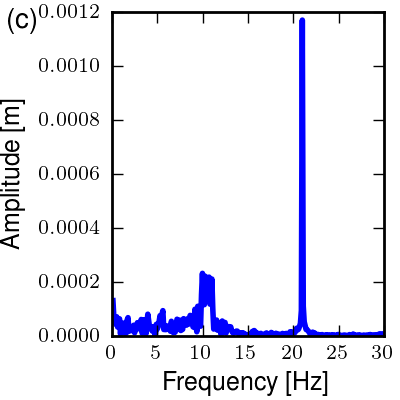}
\end{center}
\caption{Granular bed response at $f=21.0$ Hz. \textbf{(a)} Displacement of particles against the enclosure (Colors as in Fig.~\ref{fig:fig3}(a)). \textbf{(b)} Poincar\'e map. \textbf{(c)} FFT of the trajectory of the center of the mass.}
\label{fig:fig7}
\end{figure}

\begin{figure}[htp]
\begin{center}
\includegraphics[width=6cm, clip]{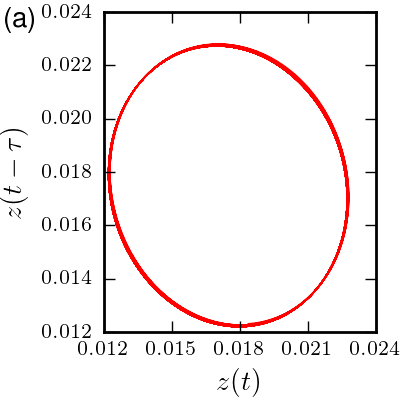}
\includegraphics[width=6cm, clip]{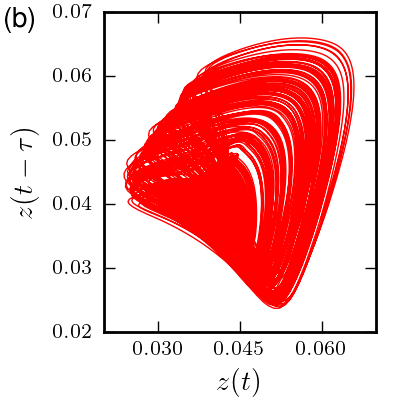}
\includegraphics[width=6cm, clip]{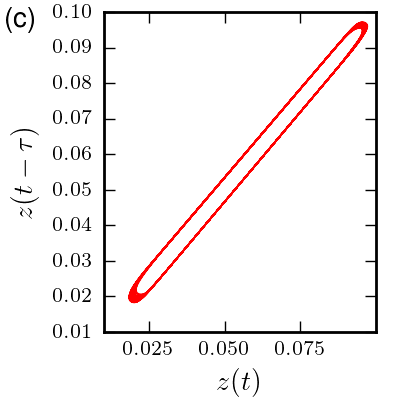}
\includegraphics[width=6cm, clip]{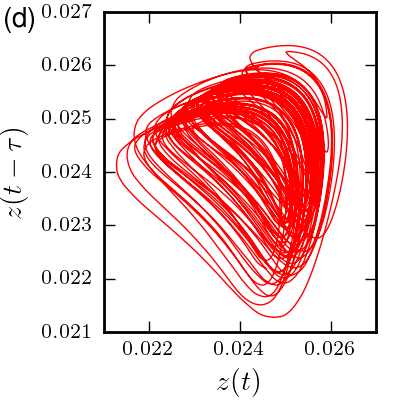}
\end{center}
\caption{State-space reconstruction. \textbf{(a)} $f=5.5$ Hz. \textbf{(b)} $f=13.0$ Hz. \textbf{(c)} $f=14.5$ Hz. \textbf{(d)} $f=21.0$ Hz.}
\label{fig:fig8}
\end{figure}

Fig.~\ref{fig:fig7} shows the above resonance chaotic behavior of the trajectory of the center of the mass of the granular bed for $f=21.0$ Hz. The cloud of points in the Poincar\'e section (Fig.~\ref{fig:fig7}(b)) and the broadband frequency observed in the power spectrum plot (Fig.~\ref{fig:fig7}(c)) suggest that the response of the grains is chaotic in this regimen. Moreover, a strange attractor is clearly identified from the reconstruction plot in (Fig.~\ref{fig:fig8}(d)).

Even though the system response presents a chaotic-like characteristic, to assure this conclusion we plot in Fig.\ref{fg:frf} the MLE for this range of frequencies.

Finally, a comparison of the motion of the granular bed in the chaotic regions show that the particles behavior is different for one or other zone.  The chaotic motion below resonance is characterized by particles colliding against the floor and ceiling. For forcing frequencies above resonance, the inertia of the granular bed does not allow the center of mass follows the motion of the box, thus only a single impact motion with the floor exists.

\section{Summary and conclusions}
\label{concl}
In this paper we study a single degree of freedom mechanical system with a prismatic PD and we have used the optimal gap size for the best performance of granular damping. In particular, we study the dynamical behavior of the center of mass of the granular bed.

We have observed that the grains run from a periodic motion to a chaotic motion when the excitation frequency increases. For low excitation frequencies the particle motion is periodic and the granular damping is very small, due to the lack of relative motion between the particles and between the particles and the enclosure. 

In contrast, for high excitation frequencies the granular bed has great energy and the movement of its center of mass behaves chaotically. Much larger frequencies lead the granular bed to expand significantly. The granular sample enters a gas-like state with only a few particles colliding with the enclosure in each oscillation. This transfers little momentum to the primary mass and the system presents low damping performance. 

Is unclear what is the correct transition to chaos, but there are signs of transition from quasi-periodicity. The emergence of the chaotic attractor in the response is preceded by the alternate appearance of quasi-periodic and periodic sub-synchronous vibrations. However, further studies are necessary to analyze intermediate frequencies to achieve a smoother transition to chaos. We found that at frequencies near the resonance of the primary system there is a window of periodicity in which the maximum granular damping is achieved. This is mainly due to the fact that the particles hit the enclosure out of phase and results in a strong reduction of the maximum displacement of the primary system.

We have found that the two chaotic zones have different behavior. Close to the resonant frequency, the particles collide against the floor and the ceiling. When the excitation frequency increases, the inertial forces do not allow the complete motion of the particles. Thus, the granular bed collides only against the floor.

\section*{Acknowledgments}
We wish to thank Diego Maza and Luis Pugnaloni, who have made valuable contributions and revisions to this work.




\bibliographystyle{elsarticle-num}




\end{document}